\documentclass[aps,prb,twocolumn,groupedaddress,10pt]{revtex4-1}
\usepackage{graphicx}
\usepackage{bm}
\usepackage{ulem}
\usepackage{slashed}
\usepackage{verbatim}
\usepackage{braket} 
\usepackage{longtable}
\usepackage{amsmath}
\usepackage{xcolor}
\usepackage{amssymb}
\usepackage{units}
\usepackage{appendix}
\usepackage[colorlinks=true,citecolor=blue,urlcolor=blue,linkcolor=black]{hyperref}
\usepackage{overpic}
\usepackage{bbold}
\usepackage{epstopdf}

\usepackage{tikz}
\usetikzlibrary{decorations.pathmorphing}
\usetikzlibrary{decorations.markings}
\usepackage{overpic}
\usetikzlibrary{patterns}
\usetikzlibrary{shapes,snakes}
\usetikzlibrary{arrows}

\begin{document}
\title{Coherent Long-Range Thermoelectrics in Nonadiabatic Driven Quantum Systems}



\author{F. Gallego-Marcos}
\email{fgallegomarcos@csic.es}
\affiliation{Instituto de Ciencia de Materiales, CSIC, Cantoblanco, 28049 Madrid, Spain}

\author{G. Platero}
\affiliation{Instituto de Ciencia de Materiales, CSIC, Cantoblanco, 28049 Madrid, Spain}


\date{\today}

\begin{abstract}
We investigate direct energy and heat transfer between two distant sites of a triple quantum dot connected to reservoirs, where one of the edge dots is driven by an ac-gate voltage. We theoretically propose how to implement heat and cooling engines mediated by long-range photoassisted transport. 
Additionally, we propose a simple set up to heat up coherently the two reservoirs symmetrically and a mechanism to store energy in the closed system. 
The present proposals can be experimentally implemented and easily controlled by tuning the external parameters.
\end{abstract}


\maketitle

%
%
%
\section{Introduction}

Quantum thermoelectric transport in nanoscale devices has gained importance due to the needs of current technology \cite{Jezouin601}. Quantum dots (QDs) have been shown to be perfect platforms to study quantum thermoelectric properties, which allow the design of thermoelectric engines \cite{PhysRevB.46.9667,PhysRevLett.89.116801,2015NatNa..10..854T,1993EL.....22...57S,1993SSCom..87.1145D,PhysRevE.81.041106,PhysRevB.87.115404}, refrigerators \cite{PhysRevLett.102.146602,PhysRevB.76.085337}, and heat rectifiers \cite{PhysRevB.83.241404,PhysRevLett.110.026804}. ac-driven thermoelectric transport  has been investigated recently, but mainly in the adiabatic regime \cite{Arrachea2007,Juergens2013,2016PhRvB..94c5436L,Ludovico2014,LudovicoOtro,2016PhRvE..93f2118B}.
\\
Recently,  experimental evidence and theoretical works show direct charge transfer between edges in arrays of QDs  by means of quantum superpositions \cite{2013NatNa...8..261B,2014PhRvL.112q6803S,2013NatNa...8..432B,2014PhRvB..89p1402S}. The mechanism behind this is termed long range (LR) transfer. 
%
%
One open question which has not yet been  addressed is if  LR energy and heat transfer could be achieved in quantum dot arrays, which are quantum simulators of real atoms and molecules.
\\
In this work we present a detailed analysis of coherent LR energy  and heat transfer in a triple quantum dot (TQD) \cite{Rogge2009,2010PhRvB..82g5304G} driven by a fast oscillating field, where we observe genuine properties of thermoelectric transport which are attributed to coherent effects. A nonadiabatic driving with frequency $\omega$ induces photon-assisted transitions (PAT) between nonresonant states detuned by $n\hbar\omega$ \cite{2013NatNa...8..432B,Gallego-marcos2014,PhysRevB.92.075302,Gallego-Marcos2016,jap..Wang..2016,1996EL.....34...43I,1997PhRvB..5512860A,2015NJPh...17e5002B,2010PhRvA..81b2117H,PhysRevB.76.085337}. We propose the PAT between LR states detuned $n\hbar\omega$ as the quantum paths to transfer coherently a controlled amount of energy between them, with only the virtual participation of the intermediate region. A mechanism to store energy in one of the quantum dots is also proposed. Furthermore, when  the outer dots are coupled to leads, we propose that these systems could work as heat and cooling engines whose  transfer mechanism is based in  photoassisted quantum superpositions between the edges. Then, a cooling engine, that we term an LR cooling engine,  works by transferring heat directly from the cold lead attached to the left dot to the hot lead attached to the right dot. A LR heat engine transfers  charge directly from the source to the drain dot against chemical potential bias. We demonstrate as well a way to symmetrically transfer energy to both leads at zero bias voltage.

\begin{figure}[b]
\centering\includegraphics[width=1.0\linewidth] {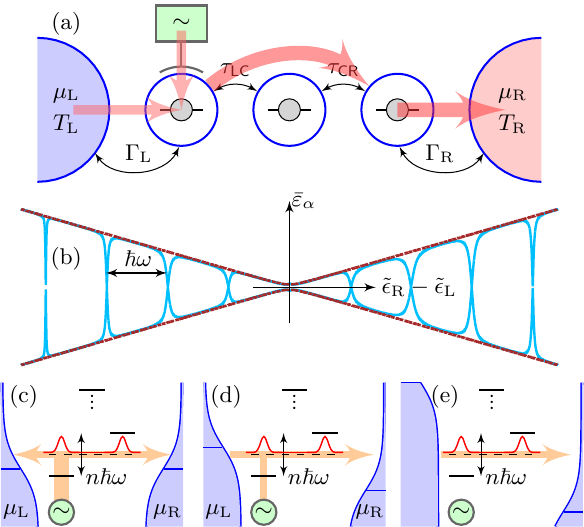}
\caption{(a) A linear TQD where the left dot is driven with an ac voltage. The thick arrows show long-range energy transfer. The energy coming from the ac  voltage and dc source is transferred directly to the right dot and then to the right contact. (b) Average value of the eigenenergies [see Eq. \eqref{app::eq::EigenVal}] vs detuning ($\Delta$) between $\ket{\text{L}}$ and $\ket{\text{R}}$ states. For zero ac driving (red dashed lines) there is a single anticrossing of the left and right levels, while for finite driving (solid blue lines) there are anticrossings at $\Delta=n\hbar\omega$ coming from the absorption or emission of $n$ photons. These anticrossings are responsible for long-range energy transfer. (c,d,e) Average energy current direction in one period of the ac field (orange arrows) for zero (c), finite (d), and infinite (e) bias voltage.}
\label{fig::esquema}
\end{figure}

%
%
%
%
%
%
\section{Driven triple quantum dot}\label{Sec::DTQD}
For simplicity we consider up to one electron in the TQD system; hence, the Hamiltonian reads: $\text{H}_\text{TQD}=\sum_{i=\{\text{L,C,R}\}}\epsilon_i\hat{\text{c}}^\dag_i\hat{\text{c}}_i+\tau_\text{LC}\hat{\text{c}}^\dag_\text{L}\hat{\text{c}}_\text{C}+\tau_\text{CR}\hat{\text{c}}^\dag_\text{C}\hat{\text{c}}_\text{R}+h.c.$, which is written in the on-site orthonormal basis:  $\ket{\text{L}}\equiv\ket{1,0,0}$, $\ket{\text{C}}\equiv\ket{0,1,0}$, $\ket{\text{R}}\equiv\ket{0,0,1}$. $\hat{\text{c}}_i$ is the fermionic destructive operator, which fulfills $\{\hat{\text{c}}_i,\hat{\text{c}}^\dag_j\}=\delta_{ij}$. The left dot is attached to an ac-gate voltage; therefore, the state $\ket{\text{L}}$ has an additional energy term, $\text{H}_\text{ac}(t)=\text{V}\sin(\omega t)\hat{\text{c}}^\dag_\text{L}\hat{\text{c}}_\text{L}$. Accordingly, the total Hamiltonian of the TQD system reads $\text{H}_\text{S}(t)=\text{H}_\text{TQD}+\text{H}_\text{ac}(t)$.
\\
We consider a configuration where the energy difference of the central dot with the outer dots is the largest energy scale in the system, i.e., $\{\text{V},\hbar\omega,|\tau_{ij}|,|\epsilon_\text{R}-\epsilon_\text{L}|\}\ll\{|\epsilon_\text{C}-\epsilon_\text{L}|,|\epsilon_\text{C}-\epsilon_\text{R}|\}$. In this regime, where the central dot is largely detuned, we develope an effective Hamiltonian by means of a perturbative scheme in the tunneling rates. The second-order effective Hamiltonian, which directly couples the outer dots by means of virtual transitions through the central region, reads \cite{Gallego-marcos2014} $\text{H}_\text{S}^\text{eff}(t)=\sum_{i=\{\text{L,R}\}}\tilde{\epsilon}_i\hat{\text{c}}^\dag_i\hat{\text{c}}_i+(\tau_{\text{co}}\hat{\text{c}}^\dag_\text{L}\hat{\text{c}}_{\text{R}}+h.c.)+\text{H}_\text{ac}(t)$, with $\tau_\text{co}=\tau_\text{LC}\tau_\text{CR}/(\epsilon_\text{R}-\epsilon_\text{C})$ and $\tilde{\epsilon}_i=\epsilon_i-\tau_{i\text{C}}^2/(\epsilon_\text{C}-\epsilon_\text{R})$. Its eigenstates and eigenenergies are obtained by Floquet theory (see Appendix \ref{app::Floquet}):
\begin{align}
\ket{\Psi_\alpha(t)}&=\sum_{k=-\infty}^\infty b_{\alpha\text{L}}^{k}(t)\ket{\text{L}}+b_{\alpha\text{R}}^{k}(t)\ket{\text{R}}\label{eq::EigenVec}\\
\varepsilon_\alpha(t)&=\sum_{k,l=-\infty}^\infty(q_\alpha-l\hbar\omega)\braket{\Phi_\alpha^{k}|\Phi_\alpha^{l}}\text{e}^{i\omega( l-k )t} \label{eq::EigenVal}
\end{align}
$b^k_{\alpha\nu}(t)=\braket{\nu|\Phi_\alpha^k}\text{e}^{-i(q_\alpha-\omega k)t}$ ($\nu=\{\text{L,R}\}$) is the weight of the $\nu$ on-site state in the $k$ mode $\ket{\Phi_\alpha^k}$ of the eigenstate $\alpha=\{1,2\}$, and $q_\alpha$ is the quasienergy. The time average energy of Eq. \eqref{eq::EigenVal}, $\bar{\varepsilon}_\alpha=\sum_{l=-\infty}^\infty(q_\alpha-l\hbar\omega)\braket{\Phi_\alpha^l|\Phi_\alpha^l}$, is plotted in Fig. \ref{fig::esquema}(b).
\\
In the open system, the Hamiltonians for the leads $\text{H}_\text{leads}$ and the interaction between the leads and the system $\text{H}_\text{int}$ read $\text{H}_\text{leads}=\sum_{\nu,k}\varepsilon_k\hat{\text{d}}_{k\nu}^\dag\hat{\text{d}}_{k\nu}$ ($\nu=\{\text{L,R}\}$) and $\text{H}_\text{int}=\sum_{\nu,k}\lambda \hat{\text{d}}_{k\nu}^\dag\hat{\text{c}}_\nu+h.c.$
\begin{figure}[t]
\centering\includegraphics[width=1\linewidth] {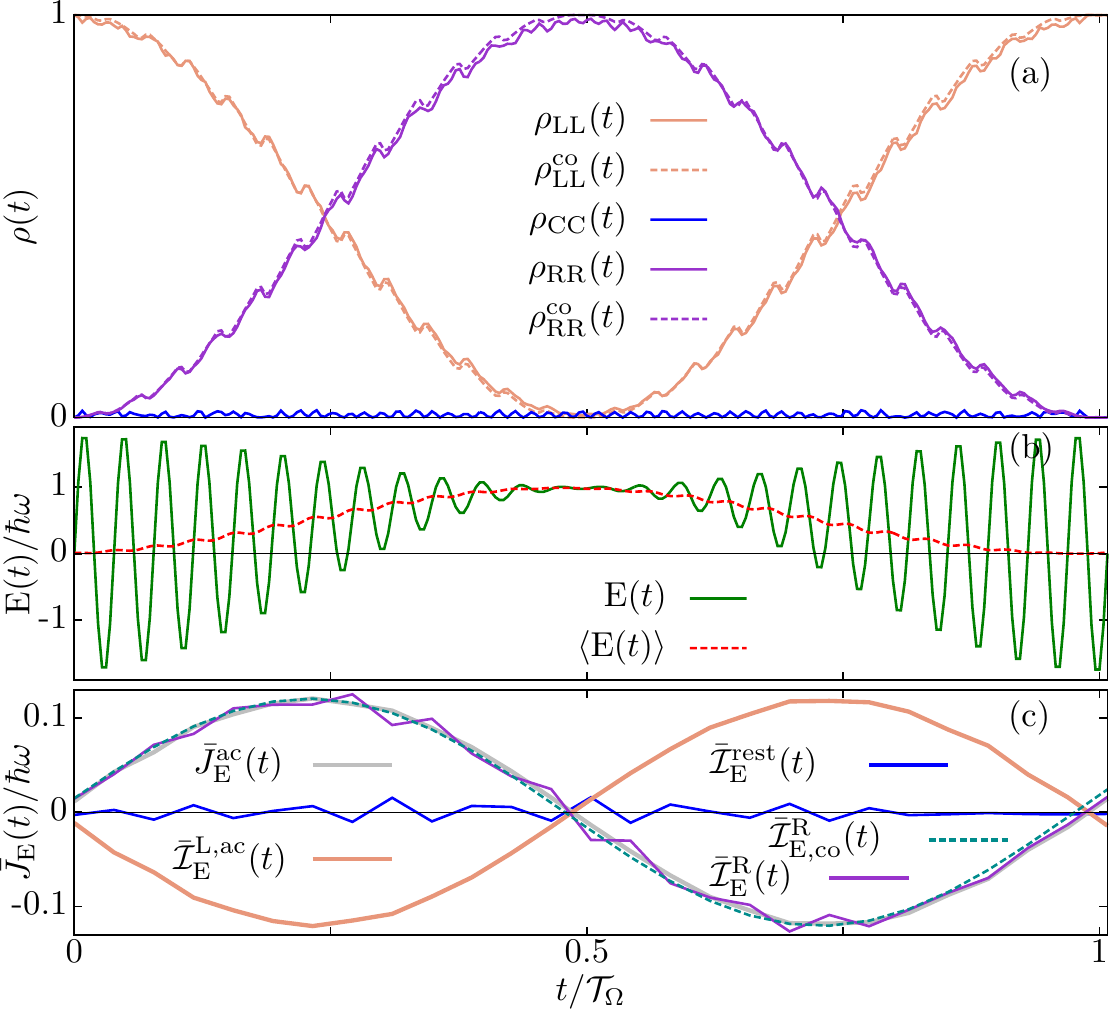}
\caption{ (a) Charge occupation evolution of the left, center and right dot for the closed system capacitively coupled to an ac gate in the cotunnel regime. The calculation is done with $\text{H}_\text{S}(t)$ (solid) and $\text{H}_\text{S}^\text{eff}(t)$ (dashed). The charge is flowing from the left to the right dot, going virtually through the central one. (b) Total energy of the TQD $\text{E}(t)$ (green line) and its average $\langle\text{E}(t)\rangle$ in a period $\mathcal{T}_\text{ac}$ (red line). (c) Average energy currents in the TQD as defined in the text. Notice that in (c) the initial value of $\bar{J}_\text{E}$  is the average value of $J_\text{E}$ in $\mathcal{T}_\text{ac}$. In (c) the current flowing to the left dot  from the ac gate [$\bar{J}_\text{E}^\text{ac}(t)$] is directly transferred to the right dot, i.e., $\bar{J}_\text{E}^\text{ac}(t)\approx \bar{\mathcal{I}}_\text{E}^\text{R}(t)$. $\hbar\omega=2\pi$, $\tau_\text{LC}=\tau_\text{CR}=\hbar\omega/2$, $\epsilon_\text{L}=0$ ($\mathcal{I}_\text{L}^{\text{L},\Omega}(t)=0$) , $\tilde{\epsilon}_\text{R}-\tilde{\epsilon}_\text{L}=\hbar\omega$, $\epsilon_\text{C}=8.5\hbar\omega$, $\text{V}=1.83\hbar\omega$, and $\mathcal{T}_\Omega=25.8\mathcal{T}_\text{ac}$.}
\label{fig::EnergyHClose}
\end{figure}

%
%
%
%

\section{Closed system}\label{sec::Close}
In an isolated TQD, the charge, energy, and entropy of the system remain constant in time. Once the ac gate is capacitively coupled to the left dot, its energy level oscillates, modifying the energy and entropy of the system.
\\
We consider the cotunnel approximation described above with $\tilde{\epsilon}_\text{L}-\tilde{\epsilon}_\text{R}=n\hbar\omega$. The density matrix $\rho(t)=\sum_\alpha p_\alpha \ket{\Psi_\alpha(t)}\bra{\Psi_\alpha(t)}$ allows one to obtain the time evolution of the system: $\dot{\rho}(t)=i/\hbar\left[\text{H}_\text{S}(t),\rho(t)\right]$.  The diagonal elements of the density matrix are plotted in Fig. \ref{fig::EnergyHClose}(a), where a direct transition between the outer dots when the central region is far detuned is shown. The energy of the TQD system $\text{E}(t)$ is determined by the expected value of the observable $\text{H}_\text{S}(t)$: $\text{E}(t)=\text{Tr}\left[\text{H}_\text{S}(t)\rho(t)\right]$. Its time evolution is related with two time scales: the ac field's period $\mathcal{T}_\text{ac}=2\pi/\omega$, and the period of the Rabi oscillations $\mathcal{T}_\Omega=2\pi/\Omega= \pi/[\text{B}_n(\text{V}/\hbar\omega)\tau_\text{co}]$, where $\text{B}_n(\alpha)$ is the n-Bessel function of first kind \cite{Gallego-marcos2014}. In the regime where $\omega\gg\Omega$, the dynamics is reflected by the average in $\mathcal{T}_\text{ac}$: 
\begin{align}
\langle\text{E}(t)\rangle\equiv \bar{\text{E}}(t)=\frac{1}{\mathcal{T}_\text{ac}}\int_t^{t+\mathcal{T}_\text{ac}}\text{E}(t')dt'. \label{eq::Average}
\end{align}
Figure \ref{fig::EnergyHClose}(b) shows the energy of the TQD during a full Rabi period: the ac gate provides energy to the system to overcome the transition to the right quantum dot and drains it to fulfill energy conservation in the right-to-left transition. The electron is initialized on $\ket{\text{L}}$ at $t=0$, and the on-site energy difference between the outer dots is $\tilde{\epsilon}_\text{R}-\tilde{\epsilon}_\text{L}=\hbar\omega$. The time evolution of the energy shows both rapid energy oscillations with period $\mathcal{T}_\text{ac}$ and slow LR Rabi oscillations; however, the main dynamics is already represented in the average energy (red line), which shows LR Rabi oscillations. 
\\
In order to study in more detail the time dependence of the energy given from the ac gate to the TQD, we calculate the time evolution of the energy:
\begin{align}
\dot{\text{E}}(t)=\text{Tr}\left[\text{H}_\text{S}(t)\dot{\rho}(t)\right]+\text{Tr}\left[\dot{\text{H}}_\text{S}(t)\rho(t)\right].\label{eq::Hder}
\end{align}
The first term in Eq. \eqref{eq::Hder} is zero in the closed system, $\text{Tr}\left[\text{H}_\text{S}(t)\dot{\rho}(t)\right]=0$, because the ac gate and the TQD system are not exchanging particles, and the TQD in the closed system is decoupled from any dissipative bath. Thus the energy current $J_\text{E}^{\text{ac}}(t)$ flowing from the ac gate to the left dot reads
\begin{align}
J_\text{E}^{\text{ac}}(t)\equiv\dot{\text{E}}(t)=\text{Tr}\left[\dot{\text{H}}_\text{ac}(t)\rho(t)\right]=\omega \text{V}\cos(\omega t)\rho_{\text{LL}}(t)\label{eq::JEac}
\end{align}
where $\rho_{\alpha\beta}(t)=\braket{\alpha|\rho(t)|\beta}$ $(\alpha,\beta=\text{L,R})$. $J_\text{E}^\text{ac}(t)$ is positive when the energy flows from the ac gate to the TQD and negative otherwise. In one oscillation of the ac driving, the energy of the charge in the left dot [$e\rho_\text{LL}(t)$] increases when $J_\text{E}^{\text{ac}}(t)>0$; on the contrary,  it decreases when $J_\text{E}^{\text{ac}}(t)<0$. If the electron occupation in the left dot remains constant during $\mathcal{T}_\text{ac}$, $\rho_\text{LL}(t)=cte$,  the averaged energy flow in one period $\mathcal{T}_\text{ac}$ will be zero: $\bar{J}_\text{E}^\text{ac}(t)=0$. But in a TQD with $\tau_{ij}\neq 0$ there are Rabi oscillations between the dots; thus $\rho_\text{LL}(t)\neq cte$ and, consequently, $\bar{J}_\text{E}^\text{ac}(t)\neq 0$.
\\
Let us focus in the internal energy current among the different regions of the TQD: $\bar{\mathcal{I}}_\text{E}^\alpha(t)$. We use the symbol $\mathcal{I}^\alpha_\text{E}(t)$ to define the energy current to the $\alpha$ region from other regions within the TQD. Their expressions are obtained from the first term on the right-hand side of Eq. \eqref{eq::Hder}, and their total sum is equal to zero $\sum_\alpha\mathcal{I}^\alpha_\text{E}=0$; see Appendix \ref{App::Internal} for their derivation. 
%
%
Due to the coupling with the ac gate, one can define the energy current to the left dot from other regions of the TQD as the sum of two different contributions: $\bar{\mathcal{I}}^\text{L}_\text{E}(t)=\bar{\mathcal{I}}_\text{E}^{\text{L},\Omega}+\bar{\mathcal{I}}_\text{E}^\text{L,ac}(t)$. 
After one oscillation, the energy level of the left dot has returned to its original value: $\epsilon_\text{L}$. The occupation and the energy in the left dot have changed to $\delta\rho_\text{LL}$ and $\epsilon_\text{L}\delta\rho_\text{LL}$, respectively. Thus $\bar{\mathcal{I}}_\text{E}^{\text{L},\Omega}\equiv \epsilon_\text{L}\delta\rho_\text{LL}/\mathcal{T}_\text{ac}$ is the energy flow considering the energy level as undriven. 
On the other hand, $\bar{\mathcal{I}}_\text{E}^\text{L,ac}(t)$ accounts for the time-resolved energy during the oscillation of the left energy level: $\bar{\mathcal{I}}_\text{E}^\text{L,ac}(t)\equiv\frac{1}{\mathcal{T}_\text{ac}}\int_{t}^{t+\mathcal{T}_\text{ac}}\text{V}\sin\omega t' \dot{\rho}_\text{LL}(t')dt'$. It actually changes the total energy of the TQD and it is equivalent to the energy given by the ac field: $\bar{J}_\text{E}^\text{ac}(t)\equiv -\bar{\mathcal{I}}_\text{E}^\text{L,ac}(t)$. 
Hence, for $\bar{J}_\text{E}^\text{ac}(t)>0$ the energy current from the ac-gate is fully transferred from the left QD to the rest of the TQD. In the cotunnel regime we have $\bar{\mathcal{I}}_\text{E}^\text{R}(t)\approx-\bar{\mathcal{I}}_\text{E}^\text{L}(t)$ [see Fig. \ref{fig::EnergyHClose}(c)], where just a negligible energy current flows to the barriers and the central dot: $\bar{\mathcal{I}}_\text{E}^\text{rest}(t)$ [see Fig. \ref{fig::EnergyHClose}(c)]. Therefore,
\begin{align}
\bar{\mathcal{I}}_\text{E}^\text{R}(t)\approx\bar{J}_\text{E}^\text{ac}(t)-\bar{\mathcal{I}}_\text{E}^{\text{L},\Omega}(t)\label{eq::JACProm}.
\end{align}
In Fig. \ref{fig::EnergyHClose}(c) we show all the internal energy currents within the TQD, $\bar{\mathcal{I}}_\text{E}(t)$, and the energy current coming from the ac gate, $\bar{J}_\text{E}^\text{ac}(t)$. We compare the current  entering the right quantum dot obtained with the full Hamiltonian, $\bar{\mathcal{I}}_\text{E}^\text{R}(t)$, with the results derived from the effective Hamiltonian, $\bar{\mathcal{I}}_\text{E,co}^\text{R}(t)$; where we find an excellent agreement.
\\
%
Initializing the system with $\rho_\text{LL}(0)=1$, in the first semiperiod $t\in\{0,\mathcal{T}_\Omega/2\}$ the energy given by the ac is stored in the right dot. In the interval, 
$t\in\{\mathcal{T}_\Omega/2,\mathcal{T}_\Omega\}$, the energy obtained from the ac gate in the previous semiperiod flows back to it: $\bar{J}_\text{E}^\text{ac}(t)<0$. A mechanism that turns off the interdot tunneling $\tau_\text{CR}$ at $t=\mathcal{T}_\Omega/2$ disconnects the right QD from the system, and the energy $\Delta\text{E}_{\mathcal{T}_\Omega/2}=\int_{0}^{\mathcal{T}_\Omega/2}\text{V}\sin(\omega t)\dot{\rho}_\text{LL}(t)dt\approx\hbar\omega$ for $\mathcal{T}_\text{ac}\ll\mathcal{T}_\Omega$ and $\text{V}\geq|\epsilon_\text{L}-\epsilon_\text{R}|$ will be stored on it. The energy can be retrieved by coupling this dot to any other system, for example, a reservoir, and transport the energy to it.
\\
The mechanism of direct energy transfer between the outer dots and the energy storage on them works also in longer QD arrays (see Appendix \ref{App::Array}). The ac driving in fact allows control of in which dot the energy is stored and the amount of it by tuning the ac voltage parameters.
\\
In order to estimate the amount of energy stored in the available experimental range, we consider $\omega=10\ \text{GHz}$, i.e., $\hbar\omega=40\mu\ \text{eV}$ and $\mathcal{T}_\text{ac}\approx 0.1\ \text{ns}$. We consider $\tau_\text{ij}$ in the range $\{0,25\}\mu\text{eV}$, and thus $\mathcal{T}_\Omega\in\{17\mathcal{T}_\text{ac},\infty\}$. Therefore, if $\tau_\text{CR}$ is set to zero at $\mathcal{T}_\Omega/2$ an amount of energy E= $40\ \mu$eV will be stored in the right dot.
\\
In summary, $\bar{J}_\text{E}^\text{ac}(t)$ directly flows from $\ket{\text{L}}$ to $\ket{\text{R}}$ and the energy can be stored in $\ket{\text{R}}$ by setting $\tau_\text{CR}=0$ at $t=\mathcal{T}_\Omega/2$.

%
%
%
%

\section{Open system}
Once the system is attached to reservoirs with chemical potentials $\mu_\nu$ and temperature $T_\nu$, it exchanges energy and charge with them. In the weak-coupling regime,  the energy shift of the on-site states due to the coupling with the reservoirs is negligible \cite{Ludovico2014}. The particle $J^{(\nu)}_\text{N}(t)$, energy $J^{(\nu)}_\text{E}(t)$, and heat $J^{(\nu)}_\text{H}(t)$ currents between the TQD and the $\nu$ reservoirs are obtained from the Born-Markov Redfield master equation (see Appendix \ref{app::Master}):
\begin{align}
J^{(\nu)}_\text{N}(t)&=\sum_{\alpha=1}^3\left[\gamma^\nu_{0\alpha}(t)\rho_\alpha(t)-\gamma^\nu_{\alpha 0}(t)\rho_0(t)\right],\label{eq::par}\\
J^{(\nu)}_\text{E}(t)&=\sum_{\alpha=1}^3\varepsilon_\alpha(t)\left[\gamma^\nu_{0\alpha}(t)\rho_\alpha(t)-\gamma^\nu_{\alpha 0}(t)\rho_0(t)\right],\label{eq::ener}\\
J^{(\nu)}_\text{H}(t)&=J^{(\nu)}_\text{E}(t)-\mu_\nu J^{(\nu)}_\text{N}(t),\quad \nu=\text{L},\text{R},\label{eq::heat}
\end{align}
where $\alpha=\{1,2,3\}$ are the subindexes for the eigenstates, the sub-index $0$ refers to the empty state, and $\gamma^\nu$ are the rates with the contacts, which read
\begin{align}
\gamma^\nu_{0\alpha}(t)&=\Gamma_\nu\sum_{\text{np}=-\infty}^\infty a_{\alpha\nu}^\text{np}(t)[1-f_{\nu}(q_\alpha-\text{p}\omega)],\\
\gamma^\nu_{\alpha0}(t)&=\Gamma_\nu\sum_{\text{ml}=-\infty}^\infty a_{\alpha\nu}^\text{ml}(t)f_{\nu}(q_\alpha-\text{l}\omega),
\end{align}
where n,p,m,l are indexes for the sidebands. $f_\nu(\varepsilon)=1/(\text{Exp}[(\varepsilon-\mu_\nu)/T_\nu]+1)$ is the Fermi function of the $\nu$ lead, $a_{\alpha\nu}^\text{np}(t)=[b_{\alpha\nu}^{\text{n}}(t)]^\dag b_{\alpha\nu}^\text{p}(t)$, and $\Gamma_\text{L}=\Gamma_\text{R}$. $b_{\alpha\nu}^\text{p}$ and $q_\alpha$ are defined in Sec. \ref{Sec::DTQD}. The currents in Eqs. \eqref{eq::par}, \eqref{eq::ener}, and \eqref{eq::heat} are defined positive when coming from the TQD system to the leads and negative otherwise. The time evolution of the energy in the TQD becomes $\dot{\text{E}}(t)=J_\text{E}^\text{ac}(t)-\sum_\nu J_\text{E}^{(\nu)}(t)$. Due to particle and energy conservation, $\bar{J}_\text{E}^{(\text{R})}=-\bar{J}_\text{E}^{(\text{L})}+\bar{J}_\text{E}^\text{ac}$ and $\bar{J}_\text{N}^{(\text{R})}=-\bar{J}_\text{N}^{(\text{L})}$.

\begin{figure}[t]
\centering\includegraphics[width=1\linewidth] {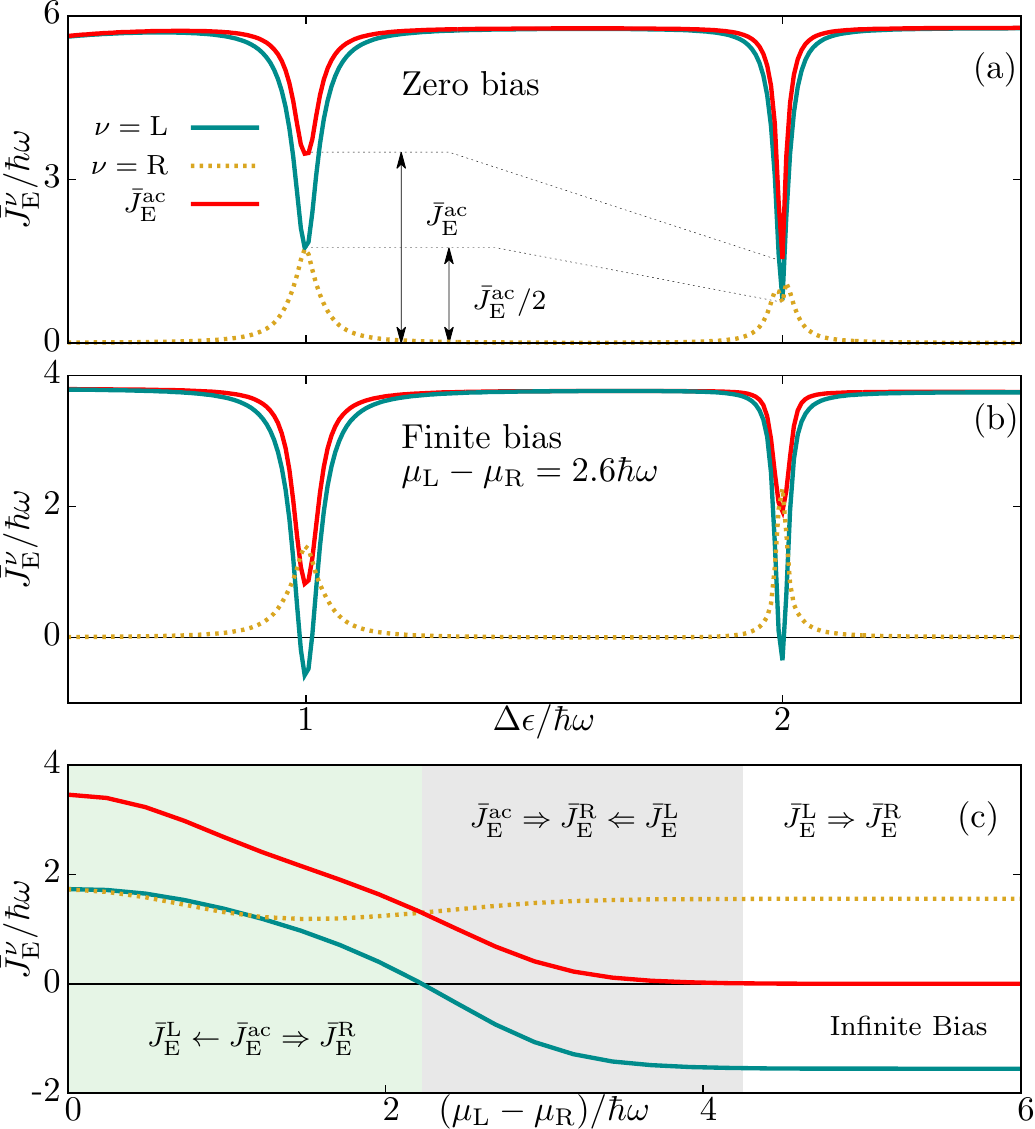}
\caption{Open system. (a,b) Average energy currents vs $\Delta\epsilon=\epsilon_\text{R}-\epsilon_\text{L}$ at zero bias (a) and finite bias (b). (c) Average energy currents vs $\mu_\text{L}-\mu_\text{R}$ for $\Delta\tilde{\epsilon}=\hbar\omega$. The arrows represent the direction of the energy current, $\rightarrow$ is a non-LR transition, and $\Rightarrow$ is a LR transition. All the energy currents are multiplied by a factor of $10^{2}$. $T_\text{L}=T_\text{R}=0.16\hbar\omega$, and the rest of the parameters are as in Fig. \ref{fig::EnergyHClose}.}
\label{fig::Zero_Inf}
\end{figure}

\begin{figure}[t]
\centering\includegraphics[width=0.7\linewidth] {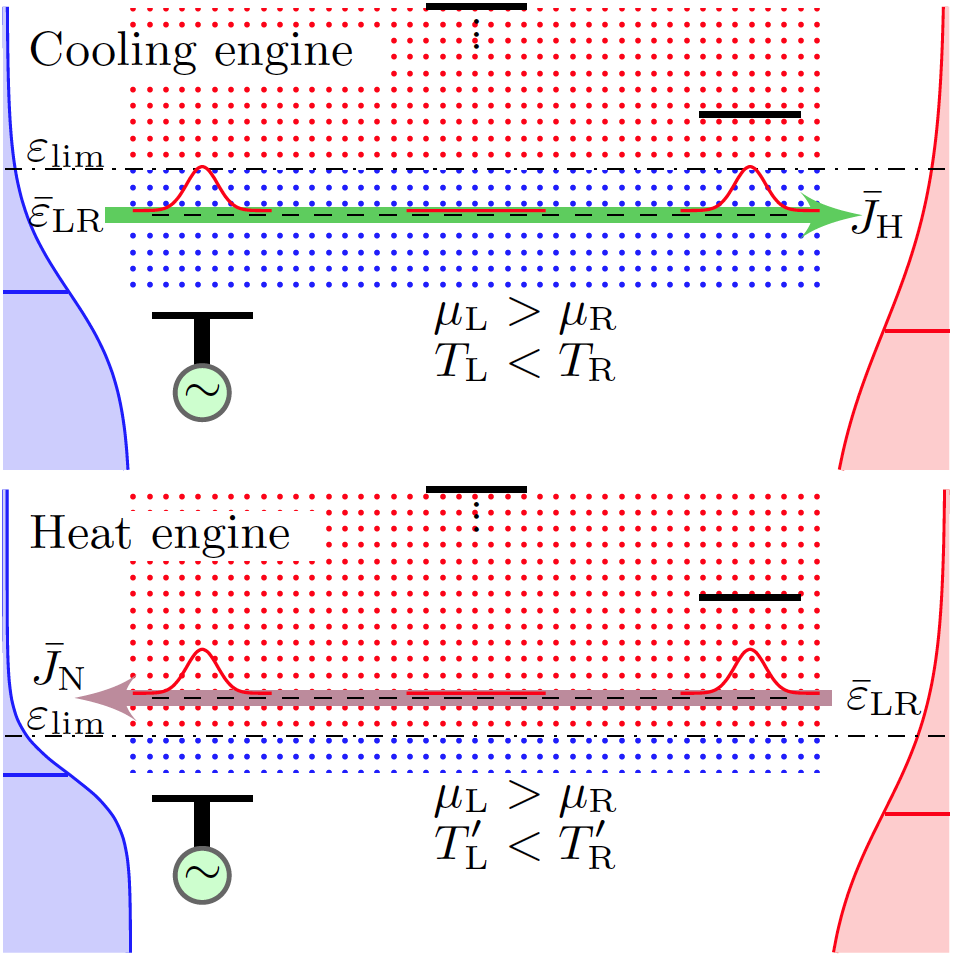}
\caption{Scheme of a cooling engine (top) and a heat engine (bottom). The blue (red) dots represent the energy area where the LR state characterizes the system as a cooling engine (heat engine), and the energy $\bar{\varepsilon}_\text{LR}$ is the average energy of the coherent states in Eq. \eqref{eq::EigenVal} for an energy difference between the outer states of $n\hbar\omega$. These two regions are limited by $\varepsilon_\text{lim}$ (dashed-dotted line). In both figures $\tilde{\epsilon}_\text{L}-\tilde{\epsilon}_\text{R}=n\hbar\omega$.}
\label{fig::ESQUEMAMaquinas}
\end{figure}

%
%
%

\subsection{Long-range energy transfer} \label{sec::LRenergy}
Our aim is to transfer energy directly to the right reservoir, with no other participation than virtual of the central region,  from the ac-gate voltage coupled to the left quantum dot and also from the left lead. It would be possible by means of the photo sidebands which participate in the LR quantum states superposition for $\tilde{\epsilon}_\text{L}-\tilde{\epsilon}_\text{R}=n\hbar\omega$ \cite{Gallego-marcos2014}.
Therefore, at $\Delta\tilde{\epsilon}=n\hbar\omega$ the system transports energy coherently among the three terminals, ac gate, and the two reservoirs through the LR superposition, going just virtually through the central dot. If the on-site energies are such that  $\Delta\tilde{\epsilon}\neq n\hbar\omega$, the two reservoirs become effectively disconnected; thus the system will behave as two separated subsystems with one reservoir coupled to a single dot. Therefore, under these conditions the energy current will just flow between the ac gate and the left reservoir $\bar{J}_\text{E}^\text{ac}=\bar{J}_\text{E}^{(\text{L})}$ while $\bar{J}_\text{E}^{(\text{R})}=0$.
\\
At zero bias $\mu_\text{L}=\mu_\text{R}$ with $\Delta\tilde{\epsilon}=n\hbar\omega$ [Fig. \ref{fig::esquema}(c)]: $f_\text{L}[\varepsilon_\alpha]=f_\text{R}[\varepsilon_\alpha]$, and the LR states are fully symmetric. Hence $J_\text{N}^{(\text{L})}(t)=J_\text{N}^{(\text{R})}(t)$ and $J_\text{E}^{(\text{L})}(t)=J_\text{E}^{(\text{R})}(t)$. Due to conservation laws $\bar{J}_\text{N}^{(\nu)}(t)=0$ but $\bar{J}_\text{E}^{(\nu)}(t)=\bar{J}_\text{E}^{\text{ac}}(t)/2$. The system produces a symmetric and continuous flow of energy from the ac gate to both reservoirs without particle flow [see Fig. \ref{fig::Zero_Inf}(a)]. 
\\
For $\mu_\text{L}>\mu_{\text{R}}$ and hence $f_\text{L}[\varepsilon_\alpha]> f_\text{R}[\varepsilon_\alpha]$, the reservoirs become asymmetric. Increasing the bias, the ac gate gives energy asymmetrically to both reservoirs until $\bar{J}_\text{E}^{(\text{L})}$ changes direction at the value where the energy current  from the left reservoir to the TQD compensates the one from the ac field to the left reservoir. Now both $\bar{J}_\text{E}^\text{ac}(t)$ and $\bar{J}_\text{E}^\text{L}$ flow directly to the right reservoir through the LR superposition, Eq. \eqref{eq::EigenVec}; see sketch in Fig. \ref{fig::esquema}(d). Figure \ref{fig::Zero_Inf}(b) shows the energy currents for finite bias. 
The ac gate keeps giving energy to the right reservoir until the bias window is sufficiently large: $f_\text{L}[\varepsilon]= 1$ and $f_\text{R}[\varepsilon]= 0$ $\forall\varepsilon$, obtaining $\rho_{ij}(t)=cte$ at the steady state and consequently, $\bar{J}_\text{E}^\text{ac}=0$ [see Eq. \eqref{eq::JEac}]. In Fig. \ref{fig::Zero_Inf}(c) the energy currents from zero to infinite bias are plotted.
\\
\ 
\\

%
%
%
%

\subsection{Long-range quantum engines} 
The construction of engines at the nanoscale is governed by the quantum properties of the  system. We define engines which work between two distant reservoirs mediated by a LR quantum superposition whose energy $\bar{\varepsilon}_\alpha$ [see Eq. \eqref{eq::EigenVal}] characterizes the type of engine. We considered engines where $\mu_\text{L}>\mu_\text{R}$ and $T_\text{L}<T_\text{R}$. We define $\varepsilon_\text{lim}$ as the energy where the system behavior changes from a heating to a cooling engine. At this energy threshold $f_\text{L}[\varepsilon_\text{lim}]=f_\text{R}[\varepsilon_\text{lim}]$. For energies $\bar{\varepsilon}_\alpha<\varepsilon_\text{lim}$, $f_\text{L}[\bar{\varepsilon}_\text{LR}]>f_\text{R}[\bar{\varepsilon}_\text{LR}]$; thus the heat flows from left to right and the system will behave as a cooling engine, transferring heat from the cold to the hot reservoir. In the other case, if $\bar{\varepsilon}_\text{LR}>\varepsilon_\text{lim}$ then $f_\text{L}[\bar{\varepsilon}_\text{LR}]<f_\text{R}[\bar{\varepsilon}_\text{LR}]$); the system behaves as a heat engine transporting particles against the chemical potential bias. In Fig. \ref{fig::ESQUEMAMaquinas} we have schematized the working region of a cooling engine (blue dots) and a heat engine (red dots). In the left figure the energy of the LR state lays in the cooling region and in the right figure in the heat region. It is important to remark that these engines, in contrast with those previously studied, work by means of virtual tunneling paths through the intermediate region of the device, i.e., the intermediated region does not participate in the transfer of heat or particles other than virtually.
\\
We are now going to consider the conditions defined above, in order to obtain a cooling engine and a heat engine in therms of the work and heat exerted on the TQD.
\\
The first law of thermodynamics for the TQD reads
\begin{align}
\dot{E}(t)=-\dot{Q}(t)-\dot{W}_\mu(t)+J_\text{E}^\text{ac}(t), \label{eq::FirsLaw}
\end{align}
where $\dot{W}_\mu(t)=\sum_\nu\mu_\nu J_\text{N}^{(\nu)}(t)$ and $\dot{Q}=\sum_\nu J_\text{H}^{(\nu)}(t)$ ($\nu=\{\text{L,R}\}$). For $\dot{W}_\mu<0$ the leads perform work on the TQD system and for $\dot{W}_\mu>0$ the TQD performs work on the leads, transporting particles against the chemical potential bias. Considering $T_\text{h}=T_\nu>T_{\nu'}=T_\text{c}$, we define $\dot{Q}^\text{h}(t)=J_\text{H}^{(\nu)}(t)$ and $\dot{Q}^\text{c}(t)=J_\text{H}^{(\nu')}(t)$ as the heat currents coming to the hot and the cold reservoirs, respectively. The second law reads \cite{2010PhRvE..82a1143E}
\begin{align}
\dot{S}(t)=\dot{S}_i(t)-\sum_\nu J_\text{H}^{(\nu)}(t)/T_\nu, \label{eq::SecondLaw}
\end{align}
where $S(t)$ is the total entropy and $\dot{S}_i(t)\geq0$ is the internal entropy production. For $\dot{S}_i(t)=0$ the system fulfills the detailed balance condition (reversible process) \cite{2010PhRvE..82a1143E}.
\\
To generate a LR quantum heat engine one needs the hot reservoir to give heat power to the TQD: $\dot{Q}^\text{h}(t)<0$. The TQD employs it to transport particles against the chemical potential bias $\dot{W}_\mu(t)>0$, charging the lead with higher chemical potential. In the LR quantum cooling engine the system cools down the cold reservoir $\dot{Q}^\text{c}(t)<0$ with the power obtained from the particle flow in the chemical potential bias direction $\dot{W}_\mu(t)<0$. The efficiencies of these two engines are
\begin{align}
\eta_\mu(t)&=\frac{\dot{W}_\mu(t)}{-\dot{Q}^\text{h}(t)}\leq \left[1-\frac{T_\text{c}}{T_\text{h}}\right]+\vartheta_\mu(t)\label{eq::effChem}\\
\eta_T(t)&=\frac{\dot{Q}^\text{c}(t)}{\dot{W}_\mu(t)}\leq\left[\frac{T_\text{c}}{T_\text{h}-T_\text{c}}\right]+\vartheta_T(t)\label{eq::effT}
\end{align}
where $\eta_\mu(t)$ ($\eta_T(t)$) is the efficiency for the heat (cooling) engine. 
\begin{figure}[t]
\centering\includegraphics[width=1\linewidth] {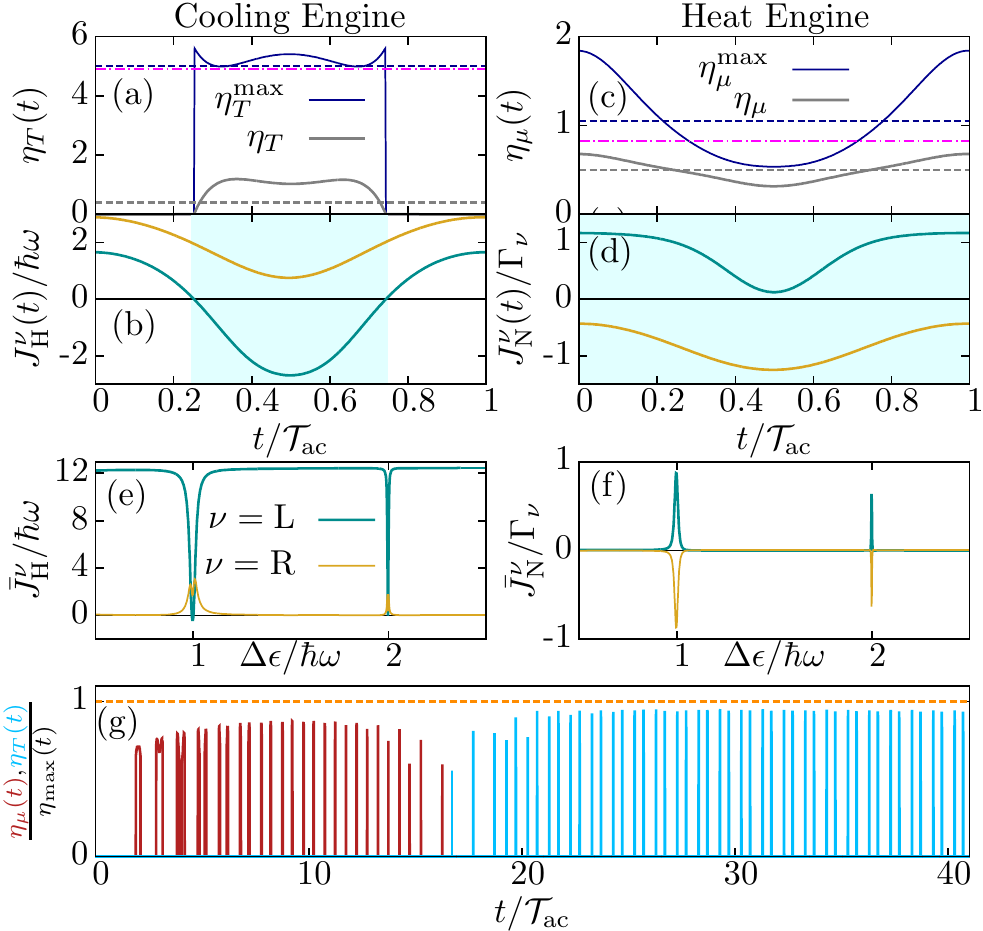}
\caption{Proposed LR cooling engine (a, b, e) with $T_\text{R}-T_\text{L}=0.03\omega$ and heat engine (c, d, f) with $T_\text{R}-T_\text{L}=0.76\omega$ in the steady state. (a, c) In gray the time-dependent efficiency  and in dark blue the maximum value of the efficiency for $\Delta\tilde{\epsilon}=\hbar\omega$. The dashed horizontal lines are the average values and the dotted-dashed pink lines are the Carnot efficiencies. (b, d, e, f) Left (right) current in dark cyan (gold). (g) Efficiency of the engines in the transitory regime with $T_\text{R}-T_\text{L}=0.13\omega$. Initially the system works as a heat engine (red) and reaches the steady state as a cooling engine (blue). $\mu_\text{L}-\mu_\text{R}=0.5\omega$, $\text{V}=0.5\omega$, and the rest of the parameters are the same as in Fig. \ref{fig::EnergyHClose}.}
\label{fig::Maquinas}
\end{figure}
The upper limits of the efficiencies are obtained from the two thermodynamic laws, Eqs. \eqref{eq::FirsLaw} and \eqref{eq::SecondLaw}, where the internal entropy production is set to zero: $\dot{S}_i=0$. In the upper limit there is a thermodynamic contribution (termed Carnot efficiencies, because they have the same expressions as the maximum efficiencies of a Carnot cycle) and a purely time-dependent term:
\begin{align}
\vartheta_\mu(t)&=\frac{\dot{E}(t)-J_\text{E}^\text{ac}(t)-\dot{S}(t)T_\text{c}}{\dot{Q}^\text{h}(t)},\ \bar{\vartheta}_\mu(t)=\frac{-\bar{J}_\text{E}^\text{ac}(t)}{\bar{\dot{Q}}^\text{h}(t)}\\
\vartheta_T(t)&=\frac{\dot{E}(t)-J_\text{E}^\text{ac}(t)-\dot{S}(t)T_\text{h}}{\dot{W}_\mu(t)},\ \bar{\vartheta}_T(t)=\frac{-\bar{J}_\text{E}^\text{ac}(t)}{\bar{\dot{W}}_\mu(t)}
\end{align}
The additional terms added to those corresponding to the Carnot efficiency in Eqs. \eqref{eq::effChem} and \eqref{eq::effT} appear in the efficiencies of heat and cooling engines out of equilibrium, where the total energy and entropy of the system depend on time \cite{2014EL....10620001A,2012AnPhy.332..110M}. In the stationary regime, in the undriven case, they are zero. But for driven systems the two variables, total energy and entropy, oscillate around the equilibrium value. Their average for each oscillation is equal to zero. The average energy current coming from the ac gate in the steady state is in general non zero (see Secs. \ref{sec::Close} and \ref{sec::LRenergy}), and for positive values it raises the maximum efficiency above the Carnot efficiency. The efficiencies obtained here should not be confused with the efficiencies of a Carnot cycle, since in the present case we are considering only engines that correspond to individual parts of the cycle. The reservoirs are continuously performing work to the TQD system for the cooling engine, which would be related to the isothermal compression of the Carnot cycle. In the case of the heat engine, the reservoirs are continuously giving heat power to the system, which would be related to the isothermal expansion of the Carnot cycle.
\\
In Fig. \ref{fig::Maquinas} (shadowed area) we plot a cooling engine (a)(b)(e) and a heat engine (c)(d)(f). These LR engines are working only when the charge is delocalized between the two outer dots of the TQD. At $\Delta\tilde{\epsilon}\neq n\hbar\omega$ the two reservoirs are disconnected; hence, as discussed in the previous section, the heat current will only flow between the ac gate and the left lead, while the right leads becomes effectively disconnected from them. In Figs. \ref{fig::Maquinas}(e) and \ref{fig::Maquinas}(f) the heat and particle currents are smaller for  $\Delta_\text{E}=2\hbar\omega$  ($n=2$) than for  $\Delta_\text{E}=\hbar\omega$  ($n=1$). The reason is that the Fermi  function at the contacts for $n=2$ is smaller than for $n=1$. Another reason is that for $n=1$ the cotunnel rate is larger for the ac field amplitude considered. Both effects are stronger for higher $n$ processes.
\\
To maximize the power and efficiencies of both, heat, and cooling engines, one should (i) select an ac field amplitude which maximizes the coupling between the outer dots at  $\Delta\tilde{\epsilon}=n\hbar\omega$, (ii) have a high-temperature bias for the heat engine, (iii) a high chemical potential bias for the cooling engine, and (iv) to operate in a high-conductance region, close to the bias window.
\\
In Fig. \ref{fig::Maquinas} the cooling engine energy region (depicted in Fig. \ref{fig::ESQUEMAMaquinas}) has an upper limit at $\varepsilon_\text{lim}=2.75\omega$ and a lower limit at $\mu_\text{L}=0.25\omega$; on the other hand, the energy region of the heat engine has a lower limit at $\varepsilon_\text{lim}=0.35\omega$. The eigenenergy of the LR superposition is $\bar{\varepsilon}_{1,2}\approx 0.9\omega$ for both cases; thus these engines are very well defined, since the energy needed to change their behavior is larger than the coupling and driving parameters of the system, $|\varepsilon_\text{lim}-\bar{\varepsilon}_{1,2}|>\{|\tau|,\text{V}\}$. However, for Fig. \ref{fig::Maquinas}(g) the energy of the LR superposition $\bar{\varepsilon}_{1,2}\approx 0.9\omega$ is close to the threshold between the two engines: $\varepsilon_\text{lim}=0.875\omega$. That is the reason why we observe that the TQD behaves as a heat engine during the transitory regime towards the steady state  before stabilizing to a cooling engine. This particular configuration, close to the threshold, is expected to be sensitive to external capacitive or thermal fluctuations.
\\
A way to experimentally observe the heat flow would be to measure the temperature or chemical potential change of the reservoirs: the bias of one thermodynamical variable of the reservoirs should be fixed either with a battery for the chemical potential or with a thermal bath for the temperature. The time evolution measurement of the other variable will give the quantity of the heat flow \cite{PhysRevA.89.013605,GallegoMarcos2014}. 
%
%
%
%
%

\section{Conclusions} 
We demonstrate  direct energy and heat transfer between outer dots without visiting (but virtually) the central site in a locally ac-driven TQD. 
Furthermore, we show how to efficiently store energy in the right dot.
As the system is attached to contacts we investigate the LR energy transport coming from  three energy sources: the two contacts attached to the TQD and the ac gate. We propose long-range quantum heat and cooling engines, driven  by high frequency, where additional tunneling channels, side bands, allow the energy and heat transfer.
\\
Our results open a way to efficiently transfer energy and heat. 
This work is easily extensible to longer arrays of  quantum dots (see Appendix \ref{App::Array}) which are within experimental reach \cite{2016NatSR...639113I}.
\\
Our device configuration is the simplest one we can consider, a single electron in a TQD, which has the ingredients to transfer heat and energy between distant sites. Our analysis can be extended to many-particle quantum dot systems, where the spin could introduce new ingredients in the behavior of  these systems  working as  quantum thermoelectrical engines. Also, interaction with other dissipative baths than the electrical contacts will be  a natural extension and it will be addressed in a future work.

%
%
%
%

\begin{acknowledgments} 
This work was supported by the Spanish Ministry of Economy and Competitiveness (MICINN) via Grant No. MAT2014-58241-P. We thank R. S\'anchez for inspiring and helpful discussions.
\end{acknowledgments}

%
%
%
%

%
%
%
%

\appendix

\begin{widetext}
\section{Internal energy currents}\label{App::Internal}
As discussed in the main text, the term $\text{Tr}\left[\text{H}_\text{S}(t)\dot{\rho}(t)\right]$ is zero:
\begin{align}
\text{Tr}\left[\text{H}_\text{S}(t)\dot{\rho}(t)\right]=i/\hbar\text{Tr}\big[\text{H}_\text{S}(t)\bigg(\text{H}_\text{S}(t)\rho(t)-\rho(t)\text{H}_\text{S}(t)\bigg)\bigg]=0.
\end{align}
However, it can be decomposed in different non zero internal TQD energy currents:
\begin{align}
\mathcal{I}_\text{E}^{\text{L}}(t)&=\mathcal{I}_\text{E}^\text{L,ac}(t)+\mathcal{I}_\text{E}^{\text{L},\Omega}(t)=\text{Tr}\left[\text{H}_\text{ac}(t)\dot{\rho}(t)\right]+\text{Tr}\left[(\epsilon_\text{L}\hat{c}^\dag_\text{L}\hat{c}_\text{L})\dot{\rho}(t)\right]\nonumber\\
&=\text{V}\sin(\omega t)\dot{\rho}_\text{LL}(t)+\epsilon_\text{L}\dot{\rho}_\text{LL}(t),\\
\mathcal{I}_\text{E}^\text{rest}(t)&=\mathcal{I}_\text{E}^\text{C}(t)+\mathcal{I}_\text{E}^\tau(t)=\text{Tr}\left[(\epsilon_\text{C}\hat{c}^\dag_\text{C}\hat{c}_\text{C})\dot{\rho}(t)\right]+\text{Tr}\left[(\tau_\text{LC}\hat{c}^\dag_\text{L}\hat{c}_\text{C}+\tau_\text{CR}\hat{c}^\dag_\text{R}\hat{c}_\text{R}+\text{H.c})\dot{\rho}(t)\right]\nonumber\\
&=\epsilon_\text{C}\dot{\rho}_\text{CC}(t)+\tau_\text{LC}\left[\dot{\rho}_\text{LC}(t)+\dot{\rho}_\text{CL}(t)\right]+\tau_\text{CR}\left[\dot{\rho}_\text{CR}(t)+\dot{\rho}_\text{RC}(t)\right],\\
\mathcal{I}_\text{E}^\text{R}(t)&=\text{Tr}\left[(\epsilon_\text{R}\hat{c}^\dag_\text{R}\hat{c}_\text{R})\dot{\rho}(t)\right]=\epsilon_\text{R}\dot{\rho}_\text{RR}(t),
\end{align}
where the superindex indicates the region of the TQD. $\mathcal{I}^\alpha_\text{E}(t)$ is positive when energy is flowing into the $\alpha$ region and negative when it is flowing out. The energy current of the left quantum dot $\mathcal{I}_\text{E}^{\text{L}}(t)$ has two terms: $\mathcal{I}_\text{E}^\text{L,ac}(t)$, which is the energy current due to the presence of the ac field, and $\mathcal{I}_\text{E}^{\text{L},\Omega}(t)$, which is the common energy flow between different regions of the TQD due to the Rabi oscillations. 

%
%
%
%

\section{Generalization to a N quantum dot array}\label{App::Array}
\begin{figure}[t]
\centering\includegraphics[width=0.45\linewidth] {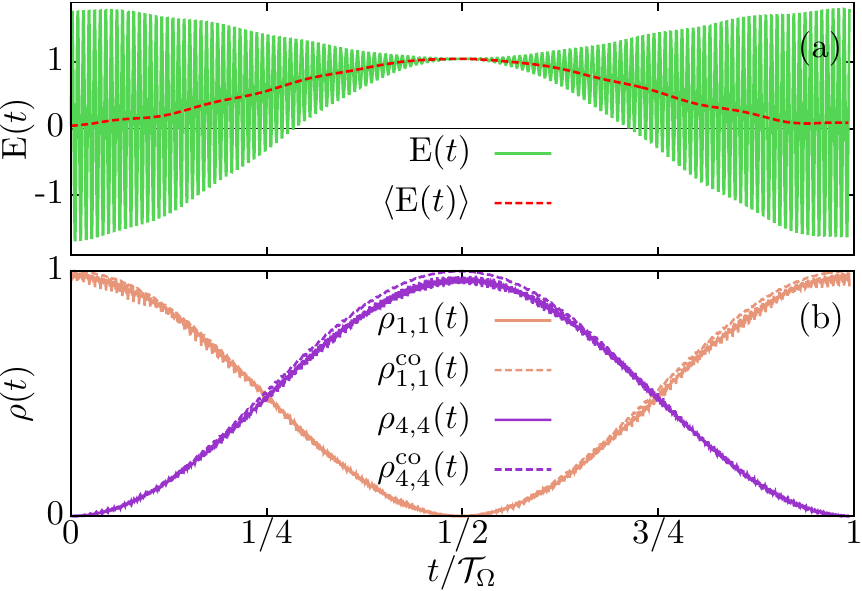}
\centering\includegraphics[width=0.45\linewidth] {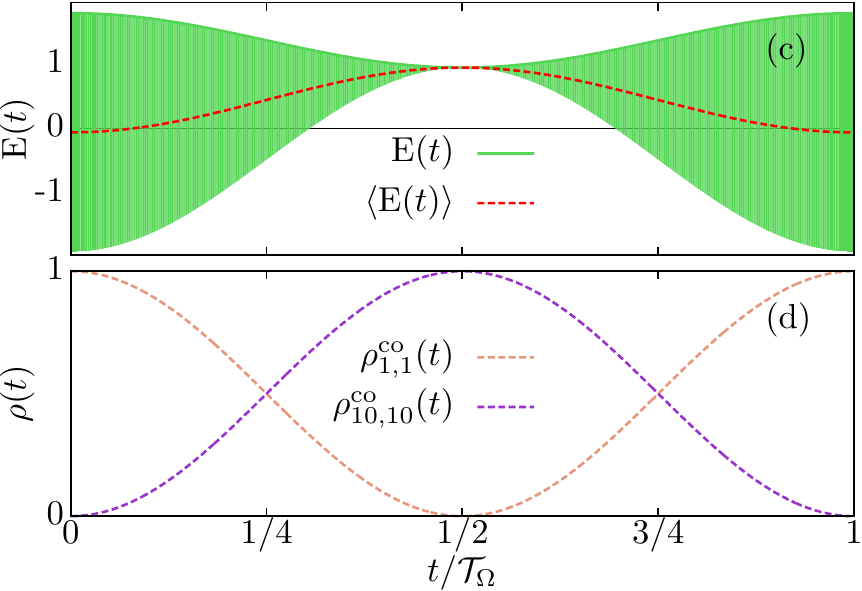}
\caption{(a) Energy and (b) charge occupation evolution for the closed system in the cotunnel regime  for $\text{N}=10$ quantum dots. $\rho(t)$ is obtained with the total Hamiltonian $\text{H}_\text{S}(t)$ and $\rho^\text{co}(t)$ with the effective Hamiltonian $\text{H}_\text{S}^\text{(\text{s}-1),eff}(t)$. We consider a detuning $\tilde{\epsilon}_4-\tilde{\epsilon}_1=\hbar\omega$, $\epsilon_{i\neq\{1,4\}}=5.5\hbar\omega$, and $\mathcal{T}^{1,4}_\Omega=139.2\mathcal{T}_\text{ac}$; hence long-range PAT transitions between dot $i=1$ and $i=4$ take place. (c, d) In this case we consider $\tilde{\epsilon}_{1}=0$, $\tilde{\epsilon}_{10}=\hbar\omega$, $\epsilon_{i\neq\{1,10\}}=5.5\hbar\omega$, and $\mathcal{T}^{1,10}_\Omega=7.39\times 10^7\mathcal{T}_\text{ac}$, obtaining long-range PAT transitions between dot $i=1$ and $i=10$. The rest of the parameters are $\hbar\omega=2\pi$, $\tau_\text{LC}=\tau_\text{CR}=\hbar\omega/2$, $\text{V}=1.83\hbar\omega$.}
\label{app::fig::Chargerho}
\end{figure}

In this section an array of N quantum dots is analyzed. The left dot ($i=1$) is capacitively coupled to an ac gate. The energy is directly transferred from the dot $i=1$ to any other dot $i=s$ whose energy level is detuned $n\hbar\omega$ with respect to the  $i=1$ quantum dot energy level, while the rest of the levels are far detuned with energy $\epsilon=\epsilon_\text{C}$. There are $\text{s}-1$ virtual states between the dot $i=1$ and $i=\text{s}$; hence to model this transition with a cotunnel Hamiltonian, we should go to the order $(\text{s}-1)$, obtaining
\begin{align}
\text{H}_\text{S}^\text{(\text{s}-1),eff}(t)=\left[\begin{array}{cc}\epsilon_\text{L}-\Lambda_\text{L}^{(\text{s}-1)}+\text{V}\text{sin}(\omega t)&-\frac{1}{(\epsilon_\text{C}-\epsilon_\text{R})^{(\text{s}-2)}}\prod_{j=1}^{\text{\text{s}-1}}\tau_{j,j+1}\\\ &\ \\-\frac{1}{(\epsilon_\text{C}-\epsilon_\text{R})^{(\text{s}-2)}}\prod_{j=1}^{\text{\text{s}-1}}\tau_{j,j+1}&\epsilon_\text{R}-\Lambda_\text{R}^{(\text{s}-1)}\\\end{array}\right]\label{app::eq::HCOT},
\end{align}
where $\Lambda_\text{i}^{(\text{s}-1)}$ is the on-site energy renormalization up to order $(\text{s}-1)$. The renormalized energies of the dots are $\tilde{\epsilon}_i=\epsilon_i-\Lambda_\text{i}^{(\text{s}-1)}$. Figures \ref{app::fig::Chargerho}(a)-\ref{app::fig::Chargerho}(d) show the results for an array of $\text{N}=10$ quantum dots. The time scale to observe a transition from the QD coupled to the ac field to the s-QD is the half period of the Rabi oscillations: $\mathcal{T}_\Omega^\text{n,s}=\pi[\epsilon_\text{C}-(\epsilon_\text{L}+n\hbar\omega)]^{\text{s}-1}/\left[\text{B}_\text{n}(\text{V}/\hbar\omega)\prod_{i=1}^{\text{s-1}}\tau_{i,i+1}\right]$, where $n$ is the sideband involved in the long-range transition. In Figs. \ref{app::fig::Chargerho}(a) and \ref{app::fig::Chargerho}(b) the energy directly flows from dot $i=1$ to $s=4$ and in Figs. \ref{app::fig::Chargerho}(c) and \ref{app::fig::Chargerho}(d) to the $\text{N}$ dot, $\text{s}=\text{N}=10$.

%
%
%

\section{Floquet Theory}\label{app::Floquet}
A time periodic Hamiltonian $\text{H}_\text{S}(t)=\text{H}_\text{S}(t+\mathcal{T}_\text{ac})$ with period $\mathcal{T}_\text{ac}=2\pi/\omega$ can be analyzed within the Floquet theory framework, which allows us to solve the evolution operator as a matrix diagonalization. The eigenvectors for the Schr\"odinger equation $\text{H}_\text{S}(t)\ket{\Psi_\alpha(t)}=i\hbar\partial_t\ket{\Psi_\alpha(t)}$ are the set $\left\{\ket{\Psi_\alpha(t)}\right\}_\alpha$. The Floquet theorem states that
\begin{align}
\ket{\Psi_\alpha(t)}=\ket{\Phi_\alpha(t)}e^{-iq_\alpha t},\quad\ \ket{\Phi_\alpha(t)}=\ket{\Phi_\alpha(t+\mathcal{T}_\text{ac})},\quad\alpha=1,2,3, \label{app::eq::FloquetTH}
\end{align}
where $q_\alpha$ are the quasienergies and $\ket{\Phi_\alpha(t)}$ are the Floquet states. Expanding $\text{H}_\text{S}(t)$ and $\ket{\Phi_\alpha(t)}$ in Fourier series $\text{H}_\text{S}(t)=\sum_{r=-\infty}^\infty\text{H}^r_\text{S} e^{ir\omega t}$ and $\ket{\Phi_\alpha(t)}=\sum_{l=-\infty}^\infty\ket{\Phi_\alpha^l} e^{il\omega t}$ the Schr\"odinger equation reads
\begin{align}
\sum_{\beta =\{\text{L,C,R}\}}\sum_{l=-\infty}^\infty\left(\text{H}^{n-l}_{\text{S},\gamma\beta}+n\omega\delta_{nl}\delta_{\gamma\beta}\right)\Phi^l_{\beta\alpha}&=q_\alpha\Phi^n_{\gamma\alpha}\label{app::eq::SolFlo}
\end{align}
where $\text{H}_{\text{S},\gamma\beta}=\braket{\gamma|\text{H}_\text{S}|\beta}$ and $\Phi^l_{\beta\alpha}=\braket{\beta|\Phi^l_{\alpha}}$, with $\{\ket{\beta}\}_\beta$ the orthonormal basis $\{\ket{\text{L}},\ket{\text{C}},\ket{\text{R}}\}$. The summation is truncated up to a value where $l\hbar\omega\gg\text{V}$. Solving Eq. \eqref{app::eq::SolFlo} we obtain from the first Brillouin zone (quasienergies $q_\alpha$ between $\{-\omega/2,\omega/2\}$) the eigenstates and the eigenenergies  of the Hamiltonian:
\begin{align}
\ket{\Psi_\alpha(t)}&=\sum_{k=-\infty}^\infty\ket{\Phi_\alpha^k}\text{e}^{-i(q_\alpha-\omega k)t}=\sum_{k=-\infty}^\infty b^k_{\alpha\text{L}}(t)\ket{\text{L}}+b^k_{\alpha\text{C}}(t)\ket{\text{C}}+b^k_{\alpha\text{R}}(t)\ket{\text{R}}\label{app::eq::EigenVec}\\
\varepsilon_\alpha(t)&=\sum_{kl=-\infty}^\infty(q_\alpha-l\hbar\omega)\braket{\Phi_\alpha^k|\Phi_\alpha^l}\text{e}^{i\omega( l-k )t},\quad \bar{\varepsilon}_\alpha=\sum_{l=-\infty}^\infty(q_\alpha-l\hbar\omega)\braket{\Phi_\alpha^l|\Phi_\alpha^l}\label{app::eq::EigenVal}
\end{align}
where $\ket{\Psi_\alpha(t)}$ is the $\alpha$ eigenstate and $\varepsilon_\alpha(t)$ the $\alpha$ eigenenergy of the TQD Hamiltonian $\text{H}_\text{S}(t)$. The expression $b^k_{\alpha\nu}(t)=\braket{\nu|\Phi_\alpha^k}\text{e}^{-i(q_\alpha-\omega k)t}$ is the weight of the $\nu$ on-site states in each sideband of the eigenstate $\alpha$. In the cotunnel approach the eigenvalues and eigenvectors of the cotunnel Hamiltonian \eqref{app::eq::HCOT} obtained with the same deviation are
\begin{align}
\ket{\Psi^\text{co}_\alpha(t)}&=\sum_k\ket{\Phi_\alpha^{k,\text{co}}}\text{e}^{-i(q^\text{co}_\alpha-\omega k)t}=\sum_{k=-\infty}^\infty b_{\alpha\text{L}}^{k,\text{co}}(t)\ket{\text{L}}+b_{\alpha\text{R}}^{k,\text{co}}(t)\ket{\text{R}}\\
\varepsilon^\text{co}_\alpha(t)&=\sum_{kl}(q^\text{co}_\alpha-l\hbar\omega)\braket{\Phi_\alpha^{k,\text{co}}|\Phi_\alpha^{l,\text{co}}}\text{e}^{i\omega( l-k )t},\quad \bar{\varepsilon}^\text{co}_\alpha=\sum_{l}(q^\text{co}_\alpha-l\hbar\omega)\braket{\Phi_\alpha^{l,\text{co}}|\Phi_\alpha^{l,\text{co}}}
\end{align}
where $\ket{\Psi^\text{co}_\alpha(t)}$ is the $\alpha$ eigenstate and $\varepsilon^\text{co}_\alpha(t)$ the $\alpha$ eigenenergy of the Hamiltonian in the cotunnel approach.
%
%
%
%

\subsection{Redfield-Master equation in Lindblad superoperator form}\label{app::Master}
For the deviation of the time evolution equation of the density matrix we make the Born-Markov approximation, where we assume that the bath correlation function decays much faster than the dynamics between the lead and the TQD, i.e., $k_\text{B}T_\nu\gg|\lambda|^2$, and that the dynamics between the TQD and the leads are much slower than the internal TQD dynamics, i.e., $|\lambda|^2\ll\tau$. We define the density matrix in the diagonalize basis (pure states) of the TQD Hamiltonian, whose eigenvalues \eqref{app::eq::EigenVal} and eigenvectors \eqref{app::eq::EigenVec} have been previously obtained with Floquet theory:
\begin{align}
\rho(t)=\sum_{\alpha=0}^3 p_\alpha\ket{\Psi_\alpha(t)}\bra{\Psi_\alpha(t)}=\sum_{\alpha=0}^3 p_\alpha\ket{\Phi_\alpha(t)}\bra{\Phi_\alpha(t)},\quad \ket{\Psi_0(t)}=\ket{\Phi_0(t)}=\ket{0}.
\end{align}
The master equation reads:
\begin{align}
\bra{\Psi_\sigma(t)}\dot{\rho}(t)\ket{\Psi_\sigma(t)}&=\sum_{\nu=\{\text{L,R}\}}\left[\sum_{\alpha=0}^3\gamma_{\sigma \alpha}^\nu(t)\bra{\Phi_\alpha(t)}\rho(t)\ket{\Phi_\alpha(t)}-\sum_{\beta=0}^3\gamma_{\beta \sigma}^\nu(t)\bra{\Phi_\sigma(t)}\rho(t)\ket{\Phi_\sigma(t)}\right]\label{app::eq::Master1}
\end{align}
where $\gamma_{\alpha\sigma}$ are the coupling rates between the leads and the TQD system. For the transition from the $\nu$ lead to $\ket{\Psi_\alpha(t)}$ ($\gamma_{\alpha0}^\nu(t)$) and for the opposite transition ($\gamma_{0\alpha}^\nu(t)$),
\begin{align}
\gamma_{0\alpha}^\nu(t)&=\sum_\text{n,p}2\pi|\lambda|^2\mathcal{D}^\nu(\Delta_{\alpha,\text{p}})\bigg\{e^{-i(\text{n}-\text{p})\omega t}\bra{\Phi^\text{n}_\alpha}\hat{c}_\nu^\dag\ket{0}\bra{0}\hat{c}_\nu\ket{\Phi^\text{p}_\alpha}\bigg\}[1-f_{\nu}(\Delta_{\alpha,\text{p}})]\equiv  \Gamma_\nu\sum_\text{np}a_{\alpha\nu}^\text{np}(t)[1-f_{\nu}(\Delta_{\alpha,\text{p}})]\label{app::eq::gama1}\\
\gamma_{\alpha 0}^\nu(t)&=\sum_\text{m,l}2\pi|\lambda|^2\mathcal{D}^\nu(\Delta_{\alpha,\text{l}})\bigg\{e^{-i(\text{l}-\text{m})\omega t}\bra{0}\hat{c}_\nu\ket{\Phi^\text{m}_\alpha}\bra{\Phi^\text{l}_\alpha}\hat{c}_\nu^\dag\ket{0}\bigg\}f_{\nu}(\Delta_{\alpha,\text{l}})\equiv  \Gamma_\nu\sum_\text{ml}a_{\alpha\nu}^\text{lm}(t)f_{\nu}(\Delta_{\alpha,\text{l}})\label{app::eq::gama2}
\end{align}
with $\Delta_{\alpha,\text{n}}=q_\alpha-\text{n}\omega$ and $\mathcal{D}^\nu(\omega)$ the density of states. We define for simplicity $\Gamma_\nu=2\pi|\lambda|^2\mathcal{D}^\nu(\varepsilon)$, which is considered constant. $a_{\alpha\nu}^\text{np}(t)=[b_{\alpha\nu}^{\text{n}}(t)]^\dag b_{\alpha\nu}^\text{p}(t)$ is the expression within the brackets. The Fermi functions $f_{\nu}(\Delta_{\alpha,\text{p}})$ are defined 
\begin{align}
f(\Delta_{\alpha,\text{p}},\mu_\nu,T_\nu)=\frac{1}{\text{e}^{(\Delta_{\alpha,\text{p}}-\mu_\nu)/k_\text{B}T_\nu}+1}\equiv f_{\nu}(\Delta_{\alpha,\text{p}})
\end{align}
which depends on the quasienergies $q_\alpha$ and on the two macroscopic values: the chemical potential $\mu_\nu$ and the temperature $T_\nu$ of the $\nu$ lead. Equation \eqref{app::eq::Master1} can be written in the form $\dot{\rho}(t)=\mathcal{L}(t)|_{\xi,\eta=0}\rho(t)$, where $\mathcal{L}(t)$ is the Liouvillian superoperator. We introduce the counting fields $\xi_\nu$ and $\eta_{\alpha\nu}$ to measure charge and energy quanta flowing through the system:
\begin{align}
\dot{\rho}(t)&=\mathcal{L}(t)|_{\xi,\eta=0}=\left.\sum_{\nu=\{\text{L,R}\}}\left[\mathcal{L}^0_\nu(t)+\sum_{\alpha=0}^3e^{i\left(\xi_\nu+\eta_{\alpha\nu}\right)}\mathcal{L}^+_{\alpha\nu}(t)+e^{-i\left(\xi_\nu+\eta_{\alpha\nu}\right)}\mathcal{L}^-_{\alpha\nu}(t)\right]\right|_{\xi,\eta=0}\rho(t)\\
\mathcal{L}^0_\nu(t)&=-\sum_{\beta\alpha=0}^3\gamma_{\beta\alpha}^\nu(t)\ket{\Psi_\alpha(t)}\bra{\Psi_\alpha(t)},\quad \mathcal{L}^+_{\alpha\nu}(t)=\gamma_{\alpha 0}^\nu(t)\ket{\Psi_\alpha(t)}\bra{\Psi_0(t)},\quad \mathcal{L}^-_{\alpha\nu}(t)=\gamma_{ 0\alpha}^\nu(t)\ket{\Psi_0(t)}\bra{\Psi_\alpha(t)}
\end{align}
The Liouvillian superoperator $\mathcal{L}(t)$ reads
\begin{align}
\mathcal{L}(t)&=\sum_{\nu=\{\text{L,R}\}}\left[\begin{array}{cccc}
-\gamma_{01}^\nu(t)&0&0&\gamma_{10}^\nu(t) \text{e}^{i(\xi_\nu+\eta_{1\nu})}\\
0&-\gamma_{02}^\nu(t)&0&\gamma_{20}^\nu(t) \text{e}^{i(\xi_\nu+\eta_{2\nu})}\\
0&0&-\gamma_{03}^\nu(t)&\gamma_{30}^\nu(t) \text{e}^{i(\xi_\nu+\eta_{3\nu})}\\
\gamma_{01}^\nu(t)\text{e}^{-i(\xi_\nu+\eta_{1\nu})}&\gamma_{02}^\nu(t)\text{e}^{-i(\xi_\nu+\eta_{2\nu})}&\gamma_{03}^\nu(t)\text{e}^{-i(\xi_\nu+\eta_{3\nu})}&-[\gamma_{10}^\nu(t)+\gamma_{20}^\nu(t)+\gamma_{30}^\nu(t)]\\\end{array}\right]\label{app::eq::MEQFB}
\end{align}

In the cotunnel approach the density matrix has one less dimension:
\begin{align}
\dot{\rho}^\text{co}(t)=\sum_{\alpha=0}^2 p_\alpha\ket{\Psi^\text{co}_\alpha(t)}\bra{\Psi^\text{co}_\alpha(t)},\quad \ket{\Psi_0(t)}=\ket{0}.
\end{align}
The Liouvillian superoperator of the master equation $\dot{\rho}^\text{co}(t)=\mathcal{L}^\text{co}(t)|_{\xi,\eta=0}\rho^\text{co}(t)$ reads
\begin{align}
\mathcal{L}^\text{co}(t)&=\sum_{\nu=\{\text{L,R}\}}\left[\begin{array}{cccc}
-\gamma_{01}^{\nu,\text{co}}(t)&0&\gamma_{10}^{\nu,\text{co}}(t) \text{e}^{i(\xi_\nu+\eta_{1\nu})}\\
0&-\gamma_{02}^{\nu,\text{co}}(t)&\gamma_{20}^{\nu,\text{co}}(t) \text{e}^{i(\xi_\nu+\eta_{2\nu})}\\
\gamma_{01}^{\nu,\text{co}}(t)\text{e}^{-i(\xi_\nu+\eta_{1\nu})}&\gamma_{02}^{\nu,\text{co}}(t)\text{e}^{-i(\xi_\nu+\eta_{2\nu})}&-[\gamma_{10}^{\nu,\text{co}}(t)+\gamma_{20}^{\nu,\text{co}}(t)]\\\end{array}\right]\label{app::eq::MEQFBCO}.
\end{align}
where $\gamma^{\nu,\text{co}}_{\alpha\beta}(t)$ has the same form as \eqref{app::eq::gama1} and \eqref{app::eq::gama2} but with one less dimension.

%
%
%
%

\subsection{Current Formulas}\label{app::Current}
With the derivatives of the counting fields $\xi$ and $\eta$ we define the particle and energy currents, respectively:
\begin{align}
J^{(\nu)}_\text{N}(t)&=\left.i\text{Tr}\left\{\partial_{\xi_\nu}\mathcal{L}(t)\rho(t)\right\}\right|_{\xi_\nu=0}\label{app::eq::par}\\
J^{(\nu)}_\text{E}(t)&=i\sum_\alpha\varepsilon_\alpha\left.\text{Tr}\left\{\partial_{\eta_{\alpha\nu}}\mathcal{L}(t)\rho(t)\right\}\right|_{\eta_{\alpha\nu}=0} \label{app::eq::ener}\\
J^{(\nu)}_\text{H}(t)&=J^{(\nu)}_\text{E}-\mu_\alpha J^{(\nu)}_\text{N}\label{app::eq::heat}
\end{align}
With some algebra the results for the currents are
\begin{align}
J^{(\nu)}_\text{N}(t)&=\text{Tr}\left\{\left[
\begin{array}{cccc}
0&0&0&-\gamma_{10}^\nu(t)\\
0&0&0&-\gamma_{20}^\nu(t)\\
0&0&0&-\gamma_{30}^\nu(t)\\
\gamma_{01}^\nu(t)&\gamma_{02}^\nu(t)&\gamma_{03}^\nu(t)&0\\\end{array}
\right]\left[
\begin{array}{c}
\rho_1(t)\\
\rho_2(t)\\
\rho_3(t)\\
\rho_0(t)\\\end{array}
\right]\right\}=\sum_{\alpha=1}^3\left[\gamma^\nu_{0\alpha}(t)\rho_\alpha(t)-\gamma^\nu_{\alpha 0}(t)\rho_0(t)\right]\label{app::eq::par2}\\
J^{(\nu)}_\text{E}(t)&=\text{Tr}\left\{\left[
\begin{array}{cccc}
0&0&0&-\varepsilon_1(t)\gamma_{10}^\nu(t)\\
0&0&0&-\varepsilon_2(t)\gamma_{20}^\nu(t)\\
0&0&0&-\varepsilon_3(t)\gamma_{30}^\nu(t)\\
\varepsilon_1(t)\gamma_{01}^\nu(t)&\varepsilon_2(t)\gamma_{02}^\nu(t)&\varepsilon_3(t)\gamma_{03}^\nu(t)&0\\\end{array}
\right]\left[
\begin{array}{c}
\rho_1(t)\\
\rho_2(t)\\
\rho_3(t)\\
\rho_0(t)\\\end{array}
\right]\right\}=\sum_{\alpha=1}^3\varepsilon_\alpha(t)\left[\gamma^\nu_{0\alpha}(t)\rho_\alpha(t)-\gamma^\nu_{\alpha 0}(t)\rho_0(t)\right]\label{app::eq::ener2}\\
J^{(\nu)}_\text{H}(t)&=\sum_{\alpha=1}^3[\varepsilon_\alpha(t)-\mu_\nu]\left[\gamma^\nu_{0\alpha}(t)\rho_\alpha(t)-\gamma^\nu_{\alpha 0}(t)\rho_0(t)\right]\label{app::eq::heat2}
\end{align}
From $\text{Tr}[\text{H}(t)\dot{\rho}(t)]$ we can also deviate the energy current:
\begin{align}
J_\text{E}^{(\nu)}(t)=\text{Tr}\left\{\left[
\begin{array}{cccc}
\varepsilon_1(t)&0&0&0\\
0&\varepsilon_2(t)&0&0\\
0&0&\varepsilon_3(t)&0\\
0&0&0&0\\\end{array}
\right]\left[
\begin{array}{c}
\dot{\rho}_1(t)=\gamma^\nu_{01}(t)\rho_1(t)-\gamma^\nu_{1 0}(t)\rho_4(t)\\
\dot{\rho}_2(t)=\gamma^\nu_{02}(t)\rho_2(t)-\gamma^\nu_{2 0}(t)\rho_4(t)\\
\dot{\rho}_3(t)=\gamma^\nu_{03}(t)\rho_3(t)-\gamma^\nu_{3 0}(t)\rho_4(t)\\
\dot{\rho}_0(t)\\\end{array}
\right]\right\}=\sum_{\alpha=1}^3\varepsilon_\alpha(t)\left[\gamma^\nu_{0\alpha}(t)\rho_\alpha(t)-\gamma^\nu_{\alpha 0}(t)\rho_0(t)\right]
\end{align}
which is exactly the same as \eqref{app::eq::ener2}.
\\
The currents for the cotunnel approach follow the same derivation as Eqs. (\ref{app::eq::par2}),(\ref{app::eq::ener2}), and (\ref{app::eq::heat2}), obtaining
\begin{align}
J^{(\nu)}_\text{N,co}(t)&=\sum_{\alpha=1}^2\left[\gamma^{\nu,\text{co}}_{0\alpha}(t)\rho_\alpha^\text{co}(t)-\gamma^{\nu,\text{co}}_{\alpha 0}(t)\rho_0^\text{co}(t)\right]\label{app::eq::parCO}\\
J^{(\nu)}_\text{E,co}(t)&=\sum_{\alpha=1}^2\varepsilon_\alpha^\text{co}(t)\left[\gamma^{\nu,\text{co}}_{0\alpha}(t)\rho_\alpha^\text{co}(t)-\gamma^{\nu,\text{co}}_{\alpha 0}(t)\rho_0^\text{co}(t)\right]\label{app::eq::enerCO}\\
J^{(\nu)}_\text{H,co}(t)&=\sum_{\alpha=1}^2[\varepsilon_\alpha^\text{co}(t)-\mu_\nu]\left[\gamma^{\nu,\text{co}}_{0\alpha}(t)\rho_\alpha^\text{co}(t)-\gamma^{\nu,\text{co}}_{\alpha 0}(t)\rho_0^\text{co}(t)\right]\label{app::eq::heatCO}
\end{align}

\end{widetext}

\bibliography{Energy_paper}
\bibliographystyle{apsrev4-1}

\end{document}